\documentclass[page-classic]{villata_2}

\title{CPT symmetry and antimatter gravity in general relativity}

\author{M. Villata\thanks{E-mail: \email{villata@oato.inaf.it}}}
\shortauthor{M. Villata}

\institute{                    
  INAF, Osservatorio Astronomico di Torino - Via Osservatorio 20, I-10025 Pino Torinese (TO), Italy
}

\abstract{
The gravitational behavior of antimatter is still unknown. While we may be confident that antimatter is self-attractive, the interaction between matter and antimatter might be either attractive or repulsive. We investigate this issue on theoretical grounds. Starting from the CPT invariance of physical laws, we transform matter into antimatter in the equations of both electrodynamics and gravitation. In the former case, the result is the well-known change of sign of the electric charge. In the latter, we find that the gravitational interaction between matter and antimatter is a mutual repulsion, i.e.\ antigravity appears as a prediction of general relativity when CPT is applied. This result supports cosmological models attempting to explain the Universe accelerated expansion in terms of a matter-antimatter repulsive interaction.}

\begin{document}

\maketitle

\section{Introduction}

The discovery of antimatter (in 1932) raised the question about its behavior in a gravitational field. Until now, no clear experimental answer could be obtained, due to the weakness of gravitation compared to the electromagnetic forces governing antiparticle motion in accelerators, and, even when dealing with electrically neutral antihydrogen, due to its fast annihilation with matter. The response will probably come in the next future from the AEGIS experiment \cite{kel08} at CERN, designed to compare the Earth gravitational acceleration on hydrogen and antihydrogen atoms.

Most physicists seem to be inclined to think that matter and antimatter must have identical gravitational properties, in the sense that gravity is always attractive. This position is supported by the fact that the physical properties distinguishing matter from antimatter (electric charge, internal quantum numbers, magnetic moment) do not affect the gravitational behavior. The obvious quantity that could make the difference, i.e.\ the mass, is requested to be positive (and equal) for both particles and antiparticles by several experimental and theoretical arguments. In particular, the main arguments against any form of gravitational repulsion (antigravity) have been those presented by Morrison \cite{mor58}, Schiff \cite{sch58,sch59}, and Good \cite{goo61}, which were later discussed and criticized by Nieto and Goldman \cite{nie91}, and further questioned by Chardin and Rax \cite{cha92,cha93,cha97}, and, in a recent paper, by Hajdukovic \cite{haj11}, so that the question is still under debate and no firm conclusion has been reached yet.

Another proof against antigravity is usually claimed to be the general theorem stating that the interaction mediated by even-spin boson fields (like the spin-2 graviton field) between like charges is attractive, whereas it would be repulsive between charges of opposite sign \cite{jag86}. Thus, in the simplistic hypothesis that the gravitational charges are the masses, since they are all positive, the interaction will be always attractive. However, as we will see in the following, in general relativity the charge is not simply the mass.

On the other hand, the idea of antigravity is as old as the discovery of antimatter, and some authors have argued on the possibility that the gravitational mass of antimatter is negative (e.g.\ \cite{haj11,noy91,ni04,haj10a,haj10b}), which would imply that matter and antimatter repel each other (but are both self-attractive). In other cases, it is proposed that antimatter is gravitationally self-repulsive (e.g.\ \cite{noy07,ben08}). All these suggestions would imply modifications to well-established theories such as general relativity.

Since the discovery of the accelerated expansion of the Universe in 1998 (e.g.\ \cite{rie98,per99}), some kind of gravitational repulsion is one of the favorite candidates to explain it, as an alternative to the dark-energy argument, and this is the reason for the current renewed interest in antigravity models (e.g.\ \cite{ni04,haj10b,noy07,ben08,rip10}).

In this paper we show that this gravitational repulsion (between matter and antimatter) can be found in the general theory of relativity without any modification to its standard formulation and with no new assumption. In particular, we assume that matter and antimatter, inertial and gravitational masses are all positive definite, together with the related energy densities, as usually requested. Moreover, we assume that general relativity is CPT invariant, even though the CPT theorem \cite{lud57}, well established in flat space-time, has not been demonstrated in curved space-time.

\section{Method and results}

Physical laws are known to be invariant under the combined CPT operations, where C (charge conjugation) is the particle-antiparticle interchange, P (parity) is the inversion of the spatial coordinates, and T is the reversal of time. Dealing with classical particles and antiparticles, it is
\begin{equation}
\label{eq.1}
{\rm CPT\,:}\qquad{\rm d}x^\mu\,\rightarrow\,-{\rm d}x^\mu\,,\qquad q\,\rightarrow\,-q\,,
\end{equation}
where $q$ is the electric charge.

This CPT symmetry implies that, if we want to transform a physical system of matter into an equivalent antimatter system (and vice versa) described by the same physical laws, it is not sufficient to replace particles with the corresponding antiparticles (C operation), but an additional PT transformation is needed too.

The PT part of eq.~(\ref{eq.1}), i.e.\ ${\rm d}x^\mu\rightarrow -{\rm d}x^\mu$, corresponds to a proper, antichronous Lorentz transformation (total inversion) represented by a diagonal matrix $(-1,-1,-1,-1)$ changing the sign of each component of any four-vector and odd-rank tensor, while it will be ineffective on even-rank tensors, being applied an even number of times. In the following, we will change the sign of any ${\rm d}x^\mu$ explicited in the equations according to eq.~(\ref{eq.1}), while the sign of the other tensors will be changed according to their rank. When CPT is applied, tensors containing the electric charge will suffer an additional sign change, so that even-rank (odd-rank) tensors, which are PT-even (PT-odd), become CPT-odd (CPT-even) in this case. The change of sign of any ${\rm d}x^\mu$ unavoidably implies that the ${\rm d}t$ of antimatter is reversed with respect to that of matter, and also the Lorentz factor $\gamma={\rm d}t/{\rm d}\tau$ will be negative for antimatter, i.e.\ while a particle has $\gamma=1$ in its rest frame, in the same reference frame an antiparticle at rest will have $\gamma=-1$.

This time inversion of antimatter is in agreement with the Feynman-St\"uckelberg interpretation \cite{stu42,fey48,fey49} that antiparticles are nothing else than the corresponding particles traveling backwards in time. Hence, when we deal with a physical system composed of both matter and antimatter, the antimatter component may be interpreted and treated as a CPT-transformed component of normal matter, since, besides applying the obvious C operation, we must reverse its time due to our observation from the opposite time direction, and the additional P operation is requested by CPT symmetry; i.e.\ as if antimatter were matter ``living" in a totally inverted space-time. Vice versa, if we want to predict the behavior of an antimatter component starting from what we know about the corresponding component of matter, we must CPT-transform the latter in the relevant equations. For example, an antiparticle (or an antimatter macroscopic body, i.e.\ made of antiparticles) in a gravitational field will behave as the CPT-transformed corresponding particle (or macroscopic matter body). Even if one does not believe in the Feynman-St\"uckelberg interpretation, CPT is in any case the only law-invariant transformation for replacing matter with antimatter in a given physical system.

When we apply CPT to all components in a certain physical law regarding matter, we obtain the corresponding behavior of antimatter, which must be the same, because of CPT invariance of that theory. In the following, we will check that all equations do not change under a whole CPT transformation.

This CPT invariance assures that antimatter is gravitationally attracted by antimatter exactly in the same way as matter is attracted by matter, but it says nothing on the interaction between matter and antimatter. And this is what we want to discover by applying the procedure outlined above.

For a useful comparison, we first see how this works in electrodynamics, then we apply to gravitation.

Throughout this paper, we use standard notations and symbols, units with $c=1$, and a $(-,+,+,+)$ metric. 

\subsection{Electrodynamics}

The equation of motion, i.e.\ the Lorentz force law, describes the dynamics of a particle of charge $q$ in an external electromagnetic field $F^{\mu\nu}$, and it is (obviously) invariant under CPT, since all the acquired minus signs cancel one another:
\begin{equation}
\label{eq.2}
{\rm CPT\,:}\qquad{{\rm d}^2x^\mu\over{\rm d}\tau^2}={q\over m}g_{\nu\lambda}F^{\mu\lambda}{{\rm d}x^\nu\over{\rm d}\tau}\quad\rightarrow\quad{-{\rm d}^2x^\mu\over{\rm d}\tau^2}={-q\over m}g_{\nu\lambda}(-F^{\mu\lambda}){-{\rm d}x^\nu\over{\rm d}\tau}\,,
\end{equation}
which means that the corresponding antimatter system obeys the same law as the original one, with both charge and field reversed in sign. For example, an electron in a proton field (hydrogen) behaves exactly as a positron in an antiproton field (antihydrogen).

If we apply CPT to the particle only (and not to the field), we have the equation for the corresponding antiparticle interacting with the original field, where one out of three acquired minus signs survives:
\begin{equation}
\label{eq.3}
{{\rm d}^2x^\mu\over{\rm d}\tau^2}=-{q\over m}g_{\nu\lambda}F^{\mu\lambda}{{\rm d}x^\nu\over{\rm d}\tau}\,,
\end{equation}
i.e.\ the equation for a particle of opposite charge, as expected (e.g., following the example above, a positron repulsed by a proton field). If we now CPT-transform all this last equation (or, which is the same thing, CPT-transform only the field in eq.~(\ref{eq.2})), we get the same equation, but now governing the motion of a normal particle in a field generated by antiparticles, namely, now the minus sign comes from the CPT-odd field and not from the charge of the particle. In any case, the net result of CPT-transforming only one of the two components, either the particle or the field, is that attraction becomes repulsion, or vice versa, since each component alone is CPT-odd.

We can further check what happens with the inhomogeneous Maxwell equations for the electromagnetic field produced by a given four-current density:
\begin{eqnarray}
\label{eq.4}
{\rm CPT\,:}\qquad{\partial\sqrt{g}F^{\mu\nu}\over\partial x^\mu}=-\sum_n{q_n\int{\delta^4(x-x_n){{\rm d}x_n^\nu\over{\rm d}\tau_n}{\rm d}\tau_n}}\quad\nonumber\\
\rightarrow\quad{\partial\sqrt{g}(-F^{\mu\nu})\over -\partial x^\mu}=-\sum_n{-q_n\int{\delta^4(x-x_n){-{\rm d}x_n^\nu\over{\rm d}\tau_n}{\rm d}\tau_n}}\,.
\end{eqnarray}
Although the charges and the field are changed in sign with respect to their matter counterparts, the equation is invariant, as it must be. In particular, we can see that the minus signs from the charges and the four-velocities cancel, so that the current density on the right-hand side is CPT-even. The field is CPT-odd, and the electromagnetic four-potential $A^\mu$, related to the field by
\begin{equation}
\label{eq.5}
F_{\mu\nu}={\partial A_\nu\over\partial x^\mu}-{\partial A_\mu\over\partial x^\nu}\,,
\end{equation}
is CPT-even.

Thus, in electrodynamics this procedure works, yielding the expected interactions for particles and antiparticles.

\subsection{Gravitation}

In the gravitational equations of the general theory of relativity, one of the most evident mathematical differences is that all tensor ranks of potentials ($g_{\mu\nu}$), fields ($\Gamma^{\lambda}_{\mu\nu}$), and currents ($T^{\mu\nu}$) are increased by one with respect to the corresponding electromagnetic quantities. This matches the fact that here the charge is no longer a scalar, but the energy-momentum four-vector $p^\mu=m\,{\rm d}x^\mu/{\rm d}\tau$. Thus, when applying CPT to gravitation, C is ineffective, but the gravitational charge changes sign by PT. Indeed, by (C)PT-transforming the energy-momentum current density $T^{\mu\nu}$,
\begin{equation}
\label{eq.6}
{1\over\sqrt{g}}\sum_n{m_n\int{\delta^4(x-x_n){{\rm d}x_n^\mu\over{\rm d}\tau_n}{{\rm d}x_n^\nu\over{\rm d}\tau_n}{\rm d}\tau_n}}\quad\rightarrow\quad{1\over\sqrt{g}}\sum_n{m_n\int{\delta^4(x-x_n){-{\rm d}x_n^\mu\over{\rm d}\tau_n}{-{\rm d}x_n^\nu\over{\rm d}\tau_n}{\rm d}\tau_n}}\,,
\end{equation}
we see that it is (C)PT-even, as expected from its rank, due to the change in sign of both the four-velocity and the charge, similarly to the electromagnetic current density in eq.~(\ref{eq.4}). Moreover, potentials ($g_{\mu\nu}$) and fields ($\Gamma^{\lambda}_{\mu\nu}$) are (C)PT-even and -odd, respectively (according to their ranks\footnote{Although the affine connection $\Gamma^{\lambda}_{\mu\nu}$ is not a tensor, its PT-oddness can be easily checked by its definition, or its relation with $g_{\mu\nu}$.}), as in electrodynamics. In summary, all quantities have the same CPT properties as their electromagnetic counterparts, because the C-oddness of the electric charge is replaced by the PT-oddness of the additional rank.

In the Einstein field equation
\begin{equation}
\label{eq.7}
R_{\mu\nu}-{1\over 2}g_{\mu\nu}R=-8\pi GT_{\mu\nu}\,,
\end{equation}
both sides are clearly even (there are only scalars and rank-2 tensors), again similarly to eq.~(\ref{eq.4}). This (C)PT invariance of the field equation implies that an antimatter energy-momentum tensor generates a gravitational field (or space-time curvature) in the same way as matter does; but, as already pointed out in electrodynamics and also in eq.~(\ref{eq.6}), with inverted charges and fields, which are both (C)PT-odd.

As a consequence of the equivalence principle, in the equation of motion (i.e.\ the geodesic equation) the mass disappears. However, in the following it may be useful to keep the ratio $m_{({\rm g})}/m_{({\rm i})}=1$ visible in the equation:
\begin{equation}
\label{eq.8}
{{\rm d}^2x^\lambda\over{\rm d}\tau^2}=-{m_{({\rm g})}\over m_{({\rm i})}}{{\rm d}x^\mu\over{\rm d}\tau}\Gamma^{\lambda}_{\mu\nu}{{\rm d}x^\nu\over{\rm d}\tau}\,.
\end{equation}
As in eq.~(\ref{eq.2}), the four-acceleration is odd, the charge $p^\mu=m_{({\rm g})}{\rm d}x^\mu/{\rm d}\tau$ is odd, as well as the field and the four-velocity, and the equation is (C)PT invariant. Therefore, an anti-apple would fall onto the head of an anti-Newton sitting on an anti-Earth, exactly in the same way as it happened here some time ago.

What about an anti-apple on the Earth, or an apple on an anti-planet? As in the electrodynamic case of eq.~(\ref{eq.2}), we must (C)PT-transform one of the two components, no matter which one, since in any case a minus sign arises, from either the (C)PT-odd field (anti-Earth) or the (C)PT-odd charge (anti-apple; in this latter case the two additional minus signs from the acceleration and velocity cancel each other). The result is a gravitational acceleration with opposite sign, i.e.\ a repulsion between matter and antimatter:
\begin{equation}
\label{eq.9}
{{\rm d}^2x^\lambda\over{\rm d}\tau^2}=-{-m_{({\rm g})}\over m_{({\rm i})}}{{\rm d}x^\mu\over{\rm d}\tau}\Gamma^{\lambda}_{\mu\nu}{{\rm d}x^\nu\over{\rm d}\tau}\,.
\end{equation}
Comparing with eq.~(\ref{eq.8}), the gravitational repulsion may be seen as the result of an effective negative gravitational mass.

\section{Discussion and conclusions}

The minus sign assigned to the gravitational mass in eq.~(\ref{eq.9}) must not be misinterpreted. It does not mean that $m_{({\rm g})}$ has become negative, since, according to our assumptions, i.e.\ CPT invariance and weak equivalence principle, all masses are and remain positive definite. As already said, the minus sign comes from the PT-oddness of either ${\rm d}x^\mu$ or $\Gamma^{\lambda}_{\mu\nu}$. Assigning it to the mass can just be useful for not losing it when dealing with the Newtonian approximation, where four-velocities disappear, together with their changed signs. Similarly, the Newtonian-limit field $GM/r^2$ has lost the PT-oddness, so that the minus sign of an antimatter field may consequently be assigned to $M$. As a result, we would obtain the generalized Newton law $F(r)=-G(\pm m)(\pm M)/r^2=\mp GmM/r^2$, where the minus sign refers to the gravitational self-attraction of both matter and antimatter, while the plus sign indicates the gravitational repulsion between matter and antimatter.

Actually, some residue of the PT properties of the geodesic equation survives in the Newton law when written in the vectorial form ${\rm d}^2{\rm\bf x}/{\rm d}t^2=-\nabla\phi$, where the P(T)-oddness of the field is in the gradient, and the triple PT-oddness of the particle is all compressed in the acceleration, with the remnant of the oddness of the charge represented by one of the two ${\rm d}t$'s at the denominator. To recognize it, it may be instructive to display the low-velocity, stationary-field approximation of the geodesic equation:
\begin{equation}
\label{eq.10}
{{\rm d}^2x^\mu\over{\rm d}\tau^2}=-{m_{({\rm g})}\over m_{({\rm i})}}{{\rm d}t\over{\rm d}\tau}\Gamma^{\mu}_{00}{{\rm d}t\over{\rm d}\tau}=-{m_{({\rm g})}\over m_{({\rm i})}}{{\rm d}t\over{\rm d}\tau}\left(-{1\over 2}g^{\mu\nu}{\partial g_{00}\over\partial x^\nu}\right){{\rm d}t\over{\rm d}\tau}\,,
\end{equation}
which is a first step towards the vectorial Newton law quoted above, achievable through the further assumption that the field is weak. In eq.~(\ref{eq.10}), the PT-odd charge $p^\mu=m_{({\rm g})}{\rm d}x^\mu/{\rm d}\tau$ has reduced to the dominant time component $p^0=E=m_{({\rm g})}{\rm d}t/{\rm d}\tau$, which, as already pointed out, is inevitably negative for antimatter due to its (P)T-oddness\footnote{However, this does not imply any negative-energy problem, since the energy density $T^{00}$ is (P)T-even and remains positive, being quadratic in ${\rm d}t/{\rm d}\tau$, see eq.~(\ref{eq.6}).}. Then, in the following scholastic passages towards the Newton law, the spatial part of eq.~(\ref{eq.10}) is divided by $({\rm d}t/{\rm d}\tau)^2$, so that the ${\rm d}t$'s disappear from the right side and appear at the denominator of the spatial acceleration on the left side, thus hiding any existence of PT-odd charges, even though the equation is still composed of two P(T)-odd components, pertaining to the particle and the field. Regarding the field, its presence as a gradient in the Newton law is already recognizable in eq.~(\ref{eq.10}).

Let us go back to the theorem quoted in the Introduction \cite{jag86}, which states that the interaction between like (opposite) charges is attractive (repulsive) when mediated by even-spin fields (like the spin-2 graviton field), and vice versa for odd-spin fields (like the spin-1 photon field). This is generally claimed to be the proof that gravity must be always attractive, since the charges, i.e.\ the masses, are always positive, even when dealing with antimatter. On the contrary, as we have stressed throughout this paper, the gravitational charge is not the mass $m_{({\rm g})}$, but $p^\mu=m_{({\rm g})}{\rm d}x^\mu/{\rm d}\tau$, which in the static case considered in the theorem reduces to the time component $p^0=m_{({\rm g})}{\rm d}t/{\rm d}\tau=m_{({\rm g})}\gamma$, with $\gamma=1$ for matter and $\gamma=-1$ for antimatter, so that gravitational repulsion between matter and antimatter is consistent with the theorem.

In conclusion, the current formulation of general relativity predicts that, while matter and antimatter are both self-attractive, matter and antimatter repel each other, under the assumption that matter is transformed into antimatter by the CPT operation expressed in eq.~(\ref{eq.1}).

This theoretical prediction of antigravity between matter and antimatter supports cosmological models attempting to explain the observed accelerated expansion of the Universe through such a repulsion between equal amounts of the two components.

The gravitational repulsion would prevent the mutual annihilation of isolated and alternated systems of matter and antimatter. The location of antimatter could be identified with the well-known large-scale (tens of Mpc) voids observed in the distribution of galaxy clusters and superclusters. Indeed, Piran \cite{pir97} showed that these voids can originate from small negative fluctuations in the primordial density field, which ({\it acting as if they have an effective negative gravitational mass}) repel surrounding matter, and grow as the largest structures in the Universe. These new cosmological scenarios could eliminate the uncomfortable presence of an unidentified dark energy, and maybe also of cosmological dark matter, which, according to the $\Lambda$-CDM concordance model, would together represent more than the 95\% of the Universe content. 

If large-scale voids are the location of antimatter, why should we not observe anything there? There is more than one possible answer, which will be investigated elsewhere.


\end{document}